\documentclass[a4paper]{IEEEtran}

\usepackage[utf8]{inputenc}
\usepackage{graphicx}
\usepackage{amssymb,amsmath,amsfonts}
\usepackage{lipsum}
\usepackage{multicol,multirow}
\usepackage{booktabs}
\usepackage[printonlyused]{acronym}
\usepackage{xcolor}
\usepackage[capitalize]{cleveref}
\crefformat{equation}{(#2#1#3)}
\crefrangeformat{equation}{(#3#1#4)-(#5#2#6)}
\crefmultiformat{equation}{(#2#1#3)}{ and~(#2#1#3)}{, (#2#1#3)}{ and~(#2#1#3)}
\usepackage{subcaption}
\usepackage{commath}
\usepackage{cite}
\usepackage[normalem]{ulem}
\usepackage{textgreek}
\usepackage{soul}
\usepackage{ulem}
\usepackage{etoolbox}
\usepackage{psfrag}
\usepackage{algorithmic}
\usepackage{textcomp}
\usepackage{algorithm}
\usepackage{graphicx}
\usepackage{epstopdf}
\usepackage[strings]{underscore}
\usepackage{caption}
\usepackage{subcaption}
\usepackage{fancyhdr}
\fancypagestyle{firstpage}{
  \fancyhf{} 
  \fancyhead[C]{\fontsize{8}{10}\selectfont
    © 2025 IEEE. Personal use of this material is permitted. \\
    This paper has been accepted for publication in the Proceedings of the IEEE 2025 Vehicular Technology Conference (VTC2025-Spring). \\
    The final version will be available via IEEE Xplore at: \texttt{[DOI to be added once available]}.}
}
\usepackage[numbers,sort&compress]{natbib}
\begin{document}

\acrodef{BS}[BS]{base station}
\acrodef{BT}[BT]{beam training}
\acrodef{CS}[CS]{compressed sensing}
\acrodef{CSI}[CSI]{channel state information}
\acrodef{CRB}[CRB]{Cramér–Rao bound}
\acrodef{CDF}[CDF]{cumulative distribution function} 
\acrodef{ECDF}[ECDF]{empirical cumulative distribution function} 
\acrodef{IO}[IO]{interacting object}
\acrodef{LS}[LS]{least-squares}
\acrodef{LOS}[LoS]{line-of-sight}
\acrodef{MMSE}[MMSE]{minimum mean squared error}
\acrodef{MIMO}[MIMO]{multiple-input multiple-output}
\acrodef{ML}[ML]{maximum likelihood}
\acrodef{MPC}[MPC]{multipath component}
\acrodef{NLOS}[NLoS]{non-line-of-sight}
\acrodef{OP}[OP]{optimal phasor}
\acrodef{RX}[RX]{receiver}
\acrodef{RIS}[RIS]{reconfigurable intelligent surface}
\acrodef{SPM}[SPM]{stationary phase method}
\acrodef{SDR}[SDR]{semi-definite relaxation}
\acrodef{SNR}[SNR]{signal-to-noise ratio}
\acrodef{TX}[TX]{transmitter}
\acrodef{TTD}[TTD]{true time delay}
\acrodef{TU}[TU]{Tunable unit}
\acrodef{ULA}[ULA]{uniform linear array}
\acrodef{UE}[UE]{user equipment}
\acrodef{UWB}[UWB]{ultra-wideband}




\title{Leveraging Large Reconfigurable Intelligent Surfaces as Anchors for Near-Field Positioning}

\author{\IEEEauthorblockN{
Zeyu Huang, \textit{Student Member, IEEE,}  
Markus Rupp, \textit{Fellow, IEEE,}   
Stefan Schwarz, \textit{Senior Member, IEEE,}
}                                     
\\
\IEEEauthorblockA{
Institute of Telecommunications, TU Wien, Vienna Austria.}  \\

E-mails:  zeyu.huang@tuwien.ac.at}

\pagestyle{empty}
\maketitle
%
\thispagestyle{firstpage}
\addtolength{\topmargin}{0.05in}

\begin{abstract}
In this work, we present a recent investigation on leveraging large \acp{RIS} as anchors for positioning in  wireless communication systems.
Unlike existing approaches, we explicitly address the uncertainty arising from the substantial physical size of the  \ac{RIS}—particularly relevant when a \ac{UE} resides in the near field—and propose a method that ensures accurate positioning under these conditions. We derive the corresponding \ac{CRB} for our scheme and validate the effectiveness of our scheme through numerical experiments, highlighting both the feasibility and potential of our approach.
\end{abstract}

\acresetall

\section{Introduction}
In future communication systems, \acp{RIS} are expected to play a pivotal role due to their ability to modify the wireless environment, traditionally viewed as static and beyond the user's control. The implementation of \acp{RIS} is projected not only to enhance communication performance but also to introduce environment-aware functionalities, thereby contributing to more adaptive and efficient systems \cite{pan2022overview,huang2022identification}.

To investigate the role of \acp{RIS} in positioning, significant efforts have been made in the literature \cite{ma2023reconfigurable}. Most \acp{RIS}-assisted positioning systems require knowledge of the \ac{RIS}-assisted channel, particularly the individual channel information between the \acp{RIS} and \acp{UE} \cite{hu2021two}. 
\ac{RIS}-assisted channel estimation typically demands considerable overhead due to the large number of elements, or pixels, on the \acp{RIS} \cite{wei2021channel,pan2022overview}. Although lower-overhead channel estimation methods have been proposed in the literature, these approaches often either compromise accuracy or introduce specific requirements that are challenging to meet.
In contrast, beam training can provide location information with much lower overhead than channel estimation \cite{wang2023hierarchical}. However, beam training is prone to beam discrimination issues due to factors such as narrow beam-widths in large arrays or the non-convex nature of RIS configuration optimization \cite{wang2023hierarchical,huang2023optimal}. When the UE is in motion or located within the near-field of the RIS \cite{lv2024ris}, the problem of beam discrimination becomes even more pronounced.

Instead of relying on \ac{RIS}-assisted channel information extraction, \acp{RIS} can serve as positioning-anchors, enabling the \ac{UE} to estimate its position independently of the specific \acp{RIS} configuration.
When \acp{RIS} are positioned far from the \ac{UE}, each \ac{RIS} can be approximated as a single point in space \cite{wang2021joint,teng2022bayesian}. However, in many scenarios, \acp{UE} are located within the near field of \acp{RIS}, particularly in millimeter-wave (mm-Wave) or higher frequency bands \cite{deng2021reconfigurable}.
The size of the RIS can also be significantly large, often referred to as an XL-RIS (extra-large RIS) \cite{yang2024near}.
In these situations, the physical size of the \ac{RIS} cannot be neglected. Considering the center of the \ac{RIS} as an anchor can also be affected by issues such as pixel failure or partial obstruction of the \ac{RIS} \cite{ozturk2024ris}.

In contrast to previous studies \cite{wang2021joint,teng2022bayesian}, we explicitly account for the inherent uncertainty introduced by the substantial dimensions of \acp{RIS} when deployed as anchors, propose a novel positioning scheme to address these effects, and derive the corresponding \ac{CRB}. Furthermore, while existing \ac{CRB} results focus on estimation-derived positioning errors, our analysis incorporates the additional uncertainty stemming from large \acp{RIS} dimensions.
The paper is organized as follows. We first introduce the system model and the positioning system in Section II. The derivation of the \ac{CRB} is elaborated in Section III. Section IV presents numerical experiments. The outlook on future work is in Section V.

\textit{Notations:} $\exp$ represents the natural exponential function, $\odot$ is the Hadamard product, $\mathrm{Diag}(\mathbf{a})$ denotes the diagonal matrix whose entries are the elements of the vector $\mathbf{a}$, $\mathrm{E} \left\{\cdot \right\}$ is the expectation operator and $[\cdot]^{\text{T}}$ is the transpose of its argument; $k, i, r$ are the index of frequency samples, time samples and \acp{RIS}, respectively, $[\cdot]_{i,j}$ means the element of the $i$-th row and $j$-th column of its argument, $\text{adj}$$[\cdot]$ means the adjugate matrix of its argument, $\sqrt{[\cdot]}$ represents the square root of each element of its argument.

\section{System Model}
\subsection{System Overview}

\begin{figure}[t!]
    \centering
    \includegraphics[width=0.85\columnwidth]{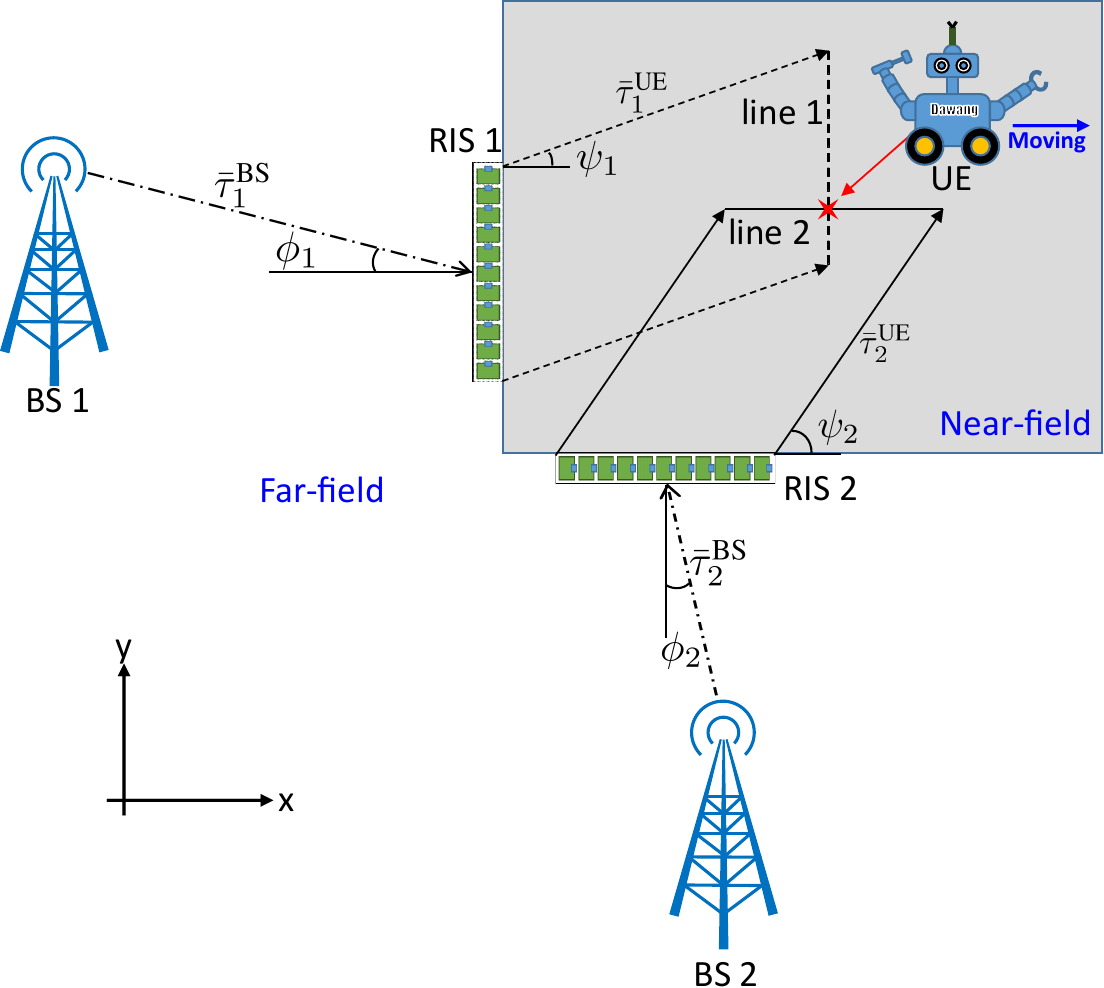}
    \caption{Multiple RISs-assisted down-link positioning system.}
    \label{fig:scenarioOverview}
\end{figure}

In this study, we investigate a \ac{RIS}-assisted downlink positioning system that consists of multiple single-antenna \acp{BS}, a moving single-antenna \ac{UE}, and multiple \acp{RIS} that relay \acp{BS} signals to the \ac{UE}, as depicted in \cref{fig:scenarioOverview}. The analysis is restricted to a two-dimensional (2D)
\footnote{The decision to use a 2D scenario is based on the following considerations. On one hand, the proposed method can be adapted to a 3D setting with modifications, but without altering the core concept. On the other hand, a simplified 2D model aids reader comprehension, and still provides insights into the proposed approach.}
scenario.  
The pixel arrays at the \acp{RIS} are modeled as \acp{ULA}. The \acp{BS} are located at a considerable distance from both the \acp{RIS} and the \ac{UE}, typically a few hundred meters away in the case of mmWave communications \cite{jirousek2021mmwave}, 
which causes the \acp{BS} to be in the far-field region of the \acp{RIS}. 
Meanwhile, the \ac{UE} is situated in the near-field region of the \acp{RIS}.
All links between the \acp{BS} and the \ac{UE} are established solely via \acp{RIS}—for example, relaying signals from rooftop to street canyon or outdoor to indoor—and a \ac{LOS} path between each \ac{RIS} and the \ac{UE} is always assumed.
 
\subsection{Signal Model}\label{sec:ChannelModel}
The received signal on pilot resources at the \ac{UE} from the \ac{BS} through the $r$-th ($r =1,\dots \mathrm{R}$) \ac{RIS} can be expressed as 
\begin{equation}\label{eq:SignalModelMatrix}
\mathbf{Y}_{r} =\mathbf{H}_{r} \odot \mathbf{X}_{r}+\mathbf{W}_{r}, \mathbf{Y}_{r}\in \mathbb{C}^{\rm{F}\times \rm{T}}
\end{equation}
where $\mathbf{H}_{r}\in \mathbb{C}^{\rm{F}\times \rm{T}}$ is the channel from the \ac{BS} through the $r$-th \ac{RIS} to the \ac{UE}, $\mathbf{X}_{r}\in \mathbb{C}^{\rm{F}\times \rm{T}}$ is the pilot signal, $\mathbf{W}_{r} \in \mathbb{C}^{\rm{F}\times \rm{T}}$ is the additive white Gaussian noise, $\mathrm{F}$ is the number of frequency samples and $\mathrm{T}$ is the number of time samples.
We define $\mathbf{h}_{r}=\mathrm{vec}\left \{ \mathbf{H}_{r} \right \}$, $\mathbf{x}_{r}=\mathrm{vec}\left \{ \mathbf{X}_{r} \right \}$, and $\mathbf{w}_{r}=\mathrm{vec}\left \{ \mathbf{W}_{r} \right \}$ and $\mathbf{y}_{r} =\mathrm{vec}\left \{ \mathbf{Y}_{r} \right \} \sim \mathcal{CN}(\mathbf{h}_{r}\odot\mathbf{x}_{r},\sigma_{\text{n}}^{2}\mathbf{I})$. 

We express the channel $\left[\mathbf{H}_{r} \right] _{k, i} $ as \cite{pan2022overview}
\begin{equation}\label{eq:ChannelModel}
\begin{split}
\left[\mathbf{H}_{r} \right] _{k, i} =  \mathbf{h}^{\text{T}}_{\text{UE},r}(f_{k},t_{i}) \mathbf{\Lambda}_{r}(f_{k},t_{i})  \mathbf{h}_{\text{BS},r}(f_{k}),
\end{split}
\end{equation}
where $f_{k}, k =1,\dots \mathrm{F}$ denotes the index of frequency samples, $t_{i}, i =1,\dots \mathrm{T}$ denotes the index of time samples, $\mathbf{\Lambda}_{r}(f_{k},t_{i})  = \text{Diag}\left(\alpha_{\text{RIS},r,1}(f_{k},t_{i}),\dots,\alpha_{\text{RIS},r,\mathrm{N}}(f_{k},t_{i})\right)$ and $\alpha_{\text{RIS},r,n}(f_{k},t_{i})\in \mathbb{C} $ is the scattering coefficient of the $n$-th \ac{RIS} pixel of the $r$-th \ac{RIS}.

The channel from the $r$-th \ac{BS} to the $r$-th \ac{RIS} $\mathbf{h}_{\text{BS},r}(f_{k})\in \mathbb {C}^{\mathrm{N}\times 1}$ is modeled using a far field channel model
\begin{equation}\label{eq:ChannelBS}
\begin{split}
\mathbf{h}_{\text{BS},r}(f_{k}) = \alpha_{\text{BS},r} \exp(-j2\pi \bar{\tau}^{\text{BS}}_r f_{k})  \boldsymbol{\alpha}_{r}(f_{k},\phi_r),
\end{split}
\end{equation}
where $\alpha_{\text{BS}}$ is the pathloss, $\bar{\tau}^{\text{BS}}_r$ is the propagation delay from the \ac{BS} to the first pixel on the $r$-th \ac{RIS}, and $\boldsymbol{\alpha}_{r}(f_{k},\phi_r)\in \mathbb {C}^{\mathrm{N}\times 1}$ is the array factor of the $r$-th \ac{RIS} w.r.t. angle of arrival $\phi_r$. The $n$-th element of $\boldsymbol{\alpha}_{r}(f_{k},\phi_r) $ can be expressed as
\begin{equation}\label{eq:ChannelBSsteering}
\begin{split}
\left[ \boldsymbol{\alpha}_{r}(f_{k},\phi_r) \right]_{n} =  \exp\left( -j2\pi f_{k} \frac{(n - 1)\Delta_\text{d}  \sin \phi } {\mathrm{c}} \right),
\end{split}
\end{equation}
where $\Delta_\text{d}$ is the pixel spacing of the \acp{RIS}.

We use $\mathbf{h}_{\text{UE},r}(f_{k},t_{i})\in \mathbb {C}^{\mathrm{N}\times 1 }$ to denote the channel from the $r$-th \ac{RIS} to the \ac{UE}.
 The $n$-th element of $\mathbf{h}_{\text{UE},r}(f_{k},t_{i}) $ can be expressed as \cite{pratschner2020measured}
\begin{equation}\label{eq:ChannelUE1}
\left [ \mathbf{h}_{\text{UE},r}(f_{k},t_{i}) \right ]_{n} = \frac{\exp\left(-j2\pi(\tau_{n,r} f_k  - \nu_{n,r} t_i)  \right)}{\text{c}\tau_{n,r} \sqrt{4\pi}},
\end{equation}
where $\tau_{n,r}$, $\nu_{n,r}$ are the delay and the Doppler frequency from the $n$-th pixel of the $r$-th \ac{RIS} to the \ac{UE}, respectively, $\text{c}$ is the speed of the light. 
We can also express \cref{eq:ChannelUE1} as
\begin{equation}\label{eq:ChannelUE2}
\begin{split}
\mathbf{h}_{\text{UE},r}(f_{k},t_{i}) = \exp\left(-j2\pi( \bar{\tau}^{\text{UE}}_r f_{k}-\bar{\nu}^{\text{UE}}_r t_{i})\right) \boldsymbol{\beta}_{r}(f_{k},t_{i}),
\end{split}
\end{equation}
where $\boldsymbol{\beta}_{r}(f_{k},t_{i}) \in \mathbb {C}^{\mathrm{N}\times 1}$ and
\begin{equation}\label{eq:ChannelUEdifference}
\begin{split}
\left[\boldsymbol{\beta}_{r}(f_{k},t_{i})\right]_{n}= \frac{\exp\left(-j2\pi (\Delta^{\tau}_{n,r} f_{k} - \Delta^{\nu}_{n,r}  t_{i})\right)}{\text{c}\tau_{n,r} \sqrt{4\pi}},
\end{split}
\end{equation}
in which 
$\bar{\tau}^{\text{UE}}_r=\frac{1}{\mathrm{N}_r} \sum_{n=1}^{\mathrm{N}_r} \tau_{n,r}$,
$\Delta^{\tau}_{n,r}=\tau_{n,r}-\bar{\tau}^{\text{UE}}_r$, $\bar{\nu}^{\text{UE}}_r=\frac{1}{\mathrm{N}_r} \sum_{n=1}^{\mathrm{N}_r} \nu_{n,r}$ and $\Delta_{\nu,n_r}=\nu_{n,r}-\bar{\nu}^{\text{UE}}_r$.


\subsection{Positioning Algorithm}\label{sec:Positioning}

\subsubsection{Estimation of propagation parameters}
Since the \acp{BS} and \acp{RIS} are pre-deployed, we assume that the channel $\mathbf{h}_{\text{BS},r}(f_{k},t_{i})$ between them is known. From \cref{eq:ChannelModel,eq:ChannelBS,eq:ChannelBSsteering,eq:ChannelUE1,eq:ChannelUE2,eq:ChannelUEdifference}, we have
\begin{equation}\label{eq:SignalModel2}
\begin{split}
\left[\mathbf{H}_{r} \right] _{k, i} \left[\mathbf{X}_{r} \right] _{k, i} =&\gamma_{r}(f_{k},t_{i}) \exp\left(-j2\pi( \bar{\tau}^{\text{UE}}_r f_{k}-\bar{\nu}^{\text{UE}}_r t_{i})\right),
\end{split}
\end{equation}
where
\begin{equation}
\begin{split}
\gamma_{r}(f_{k},t_{i}) &=\boldsymbol{\beta}^{\text{T}}_{r}(f_{k},t_{i}) \mathbf{\Lambda}_{r}(f_{k},t_{i})  \mathbf{h}_{\text{BS},r}(f_{k}) \left[\mathbf{X}_{r} \right] _{k, i}.
\end{split}
\end{equation}
The estimation of $\bar{\tau}^{\text{UE}}_r$ and $\bar{\nu}^{\text{UE}}_r$ can be achieved through various well-established parameter estimation algorithms, such as ESPRIT, MUSIC or SAGE \cite{yin2016propagation}. 
Based on $\bar{\tau}^{\text{UE}}_{r}$, $\bar{\nu}^{\text{UE}}_{r}$ 
and the direction of movement of the UE, we can determine the propagation distance $d_{r}$ of the \ac{MPC} from the $r$-th  \ac{RIS} to the \ac{UE}, as well as the angle $\psi_{r}$ between the direction of the \ac{UE}’s motion and the propagation direction of the \ac{MPC}.
Specifically, we have
\begin{equation}\label{eq:DopplerToAngle}
\begin{split}
&d_{r}=\rm{c} \bar{\tau}^{\text{UE}}_{r},\\
&\psi_{r}=\mathrm{arccos}\left(\frac{\mathrm{c}\bar{\nu}^{\text{UE}}_{r}}{ \mathrm{v} f_{k}}\right), 0<\psi_{r}\leq\pi,
\end{split}
\end{equation}
where $\mathrm{v}$ represents the speed of the \ac{UE}.
In this work, we align the x-axis with the \ac{UE}'s direction of travel and assume that the \ac{UE} knows on which side each \ac{RIS} is located.

\subsubsection{Position information from a single \ac{RIS}}
We define the point on the $r$-th \ac{RIS}, which satisfies the condition that the delay from this point to the \ac{UE} is $\bar{\tau}^{\text{UE}}_r$, and the angle between the direction of the \ac{UE}’s motion and the propagation direction of the \ac{MPC} is  $\psi_{r}$, as the $r$-th anchor point. 
To account for the uncertainty stemming from the large physical size of the \ac{RIS}, we assume that the anchor point can be located anywhere on the \ac{RIS}, corresponding to a uniform prior distribution of the anchor location\footnote{In this section, we illustrate our scheme under noise-free conditions. The impact of noise is then incorporated through the \ac{CRB} analysis in Section in \cref{sec:CRB}.}.
Consequently, the position of the UE could be any point determined by the shift of the $r$-th anchor according to $(d_{r}\mathrm{cos}\psi_{r},d_{r}\mathrm{sin}\psi_{r})$.
For instance, in \cref{fig:scenarioOverview} the \ac{UE} estimates $\bar{\tau}_{1}$ and $\psi_{1}$ for the \ac{MPC} originating from \ac{RIS}\,1, resulting in all possible locations of the \ac{UE} being represented by the dashed line labeled as line\,1.

\subsubsection{Position information from multiple \acp{RIS}}
Although a single \ac{RIS} in our setup is insufficient to determine the precise location of the \ac{UE}, combining information from another \ac{RIS} makes this feasible. This is demonstrated with \ac{RIS}\,2 in \cref{fig:scenarioOverview}. Similar to \ac{RIS}\,1, we derive another line segment, labeled as line 2, which is represented by a solid line that contains all potential position information for the \ac{UE} as derived from \ac{RIS}\,2. When \ac{RIS}\,1 and \ac{RIS}\,2 are not parallel, the intersection of line\,1 and line\,2 yields the estimated position of the \ac{UE}, which is marked with a red star.

\subsubsection{Position  estimation}
The position of the $r$-the \ac{RIS} can be characterized by
\begin{equation}\label{eq:RISposition}
\tilde{y}=k_{r} \tilde{x}+ \tilde{b_{r}}, \tilde{x} \in \tilde{\mathbb{S}}_{r},
\end{equation}
where $\tilde{\mathbb{S}}_{r}$ represents the set of x-coordinates of the $r$-th \ac{RIS}, i.e., the projection of the line-segment onto the x-axis.
Then the estimated possible position set of the \ac{UE}, derived from the $r$-th \ac{RIS} can be expressed as
\begin{equation}\label{eq:UEpositionr}
y=k_{r} x+ \tilde{b_{r}} - k_{r} d_{r} \cos \psi_{r}+d_{r} \sin \psi_{r}, x \in \mathbb{S}_{r}, 
\end{equation}
where $\{x \in \mathbb{S}_{r}| x = \tilde{x} + d_{r} \mathrm{cos}\psi_{r},\tilde{x} \in \tilde{\mathbb{S}}_{r} \}$.
Considering all the \acp{RIS}, we have a linear system of equations
\begin{equation}\label{eq:UEposition5}
\mathbf{K}\mathbf{p}=\begin{bmatrix}
1 & -k_{1}\\ 
\vdots & \vdots\\ 
1 & -k_{\text{R}}
\end{bmatrix}
\begin{bmatrix}
y\\ 
x
\end{bmatrix}
=
\begin{bmatrix}
b_{1}\\ 
\vdots\\ 
b_{\text{R}}
\end{bmatrix}=\mathbf{b}(\boldsymbol{\theta}),
\end{equation}
where $b_r = \tilde{b_{r}} - k_{r} d_{r} \cos \psi_{r}+d_{r} \sin \psi_{r}$, $\boldsymbol{\theta}=[\boldsymbol{\theta}_{1}; \dots; \boldsymbol{\theta}_{\text{R}}]$, $\boldsymbol{\theta}_{r} = [\bar{\tau}^{\text{UE}}_r,\bar{\nu}^{\text{UE}}_r]^{\text{T}}$.
When $\mathbf{K}$ is a full-rank matrix, ensuring that $\mathbf{K}^{\#}=\left(\mathbf{K}^{\text{T}}\mathbf{K}\right)^{-1} \mathbf{K}^{\text{T}}$, the pseudo-inverse of $\mathbf{K}$ exists,
the \ac{LS} estimate of the \ac{UE} position is
\begin{equation}\label{eq:UEposition7}
\hat{\mathbf{p}}=\mathbf{K}^{\#} \mathbf{b}(\boldsymbol{\theta}).
\end{equation}

\section{Performance Bound Analysis}\label{sec:CRB}
Considering the vectorized system model \cref{eq:SignalModelMatrix}, the log-likelihood function w.r.t. $\bar{\tau}^{\text{UE}}_r$ and $\bar{\nu}^{\text{UE}}_r$ can be defined as \cite{kay1993statistical}
\begin{equation}\label{eq:likelihood}
\begin{split}
& \mathrm{L}( \bar{\tau}^{\text{UE}}_r, \bar{\nu}^{\text{UE}}_r) =-\frac{1}{\sigma^{2}_{\text{n}}}(\mathbf{y}_{r}-\mathbf{h}_{r}\odot\mathbf{x}_{r})^{\text{H}}(\mathbf{y}_{r}-\mathbf{h}_{r}\odot\mathbf{x}_{r}).
\end{split}
\end{equation}
The Fisher information matrix can be written as
\begin{equation}
\mathbf{J}(\boldsymbol{\theta}_{r})=
\begin{bmatrix}
\mathbf{g}^{\text{T}}_{r} \mathbf{f} & \mathbf{g}^{\text{T}}_{r} \mathbf{k} \\ 
\mathbf{g}^{\text{T}}_{r} \mathbf{k} & \mathbf{g}^{\text{T}}_{r} \mathbf{t}
\end{bmatrix},
\end{equation}
where
\begin{equation}
\begin{split}
&\mathbf{g}_{r} =  \mathbf{h}_{r} \odot \mathbf{x}_{r} \odot \mathbf{h}^{*}_{r} \odot \mathbf{x}^{*}_{r}   \in \mathbb {R}^{\rm{FT}\times 1},\\
&\mathbf{f} = \frac{8 \pi^{2}}{\sigma^{2}_{\text{n}}} \left[ f^{2}_{1},\dots,  f^{2}_{\rm{F}},\dots, f^{2}_{1},\dots,  f^{2}_{\rm{F}}\right]^{\text{T}}\in \mathbb {R}^{\rm{FT}\times 1},\\
&\mathbf{t} = \frac{8 \pi^{2}}{\sigma^{2}_{\text{n}}} \left[ t^{2}_{1},\dots,t^{2}_{1},\dots,  t^{2}_{\rm{T}}\dots,  t^{2}_{\rm{T}} \right]^{\text{T}}\in \mathbb {R}^{\rm{FT}\times 1},\\
&\mathbf{k} = -\sqrt{\mathbf{f} \odot \mathbf{t}}\in \mathbb {R}^{\rm{FT}\times 1}.
\end{split}
\end{equation}
Based on the chain rule \cite{kay1993statistical}, we have
\begin{equation}\label{eq:FisherInformationP1}
\begin{split}
\text{cov}\{\hat{\mathbf{p}}-\mathbf{p}\} =\mathrm{E}\{\left(\hat{\mathbf{p}}-\mathbf{p}\right) \left(\hat{\mathbf{p}}-\mathbf{p}\right)^{\text{T}}\}  \geq   \mathbf{K}^{\#}  \mathbf{C} {\mathbf{K}^{\#}}^{\text{T}},\\
\end{split}
\end{equation}
where $\mathbf{p}$ is the true position of the \ac{UE},  $\mathbf{C}= \text{Diag}\left(c_{1},\dots,c_{\text{R}}\right)$ and the $r$-th element on the main diagonal is
\begin{equation}
\begin{split}
c_{r}
&= \left[\frac{\partial [\mathbf{b}(\boldsymbol{\theta})]_{r}}{\partial\bar{\tau}^{\text{UE}}_r} ,\frac{\partial [\mathbf{b}(\boldsymbol{\theta})]_{r}}{\partial\bar{\nu}^{\text{UE}}_r}  \right] \mathbf{J}^{-1}\left(\boldsymbol{\theta}_{r}\right) 
\begin{bmatrix}
\dfrac{\partial [\mathbf{b}(\boldsymbol{\theta})]_{r}}{\partial\bar{\tau}^{\text{UE}}_r}  \\ 
\dfrac{\partial [\mathbf{b}(\boldsymbol{\theta})]_{r}}{\partial\bar{\nu}^{\text{UE}}_r}  \\ 
\end{bmatrix},\\
\end{split}
\end{equation}
where
\begin{equation}\label{eq:partialDerivativesTau1}
 \frac{\partial [\mathbf{b}(\boldsymbol{\theta})]_{r}}{\partial\bar{\tau}^{\text{UE}}_r}
=-\text{c}\left( k_r \cos \psi_{r}+ \sin \psi_{r}\right),
\end{equation}
\begin{equation}\label{eq:partialDerivativesNu1}
\begin{split}
 \frac{\partial [\mathbf{b}(\boldsymbol{\theta})]_{r}}{\partial\bar{\nu}^{\text{UE}}_r}
=-\frac{\left( k_{r}\sin\psi_{r}+\cos\psi_{r}\right)\text{c}^{2}\bar{\tau}^{\text{UE}}_r}{\mathrm{v}f\sqrt{1-\left(\frac{\mathrm{c}\bar{\nu}^{\text{UE}}_r}{\mathrm{v}f}\right)^{2}}}.
\end{split}
\end{equation}
The inverse of $\mathbf{J}(\boldsymbol{\theta}_{r}) $ can be expressed as 
\begin{equation}\label{eq:InverseMatrixTheta3}
\mathbf{J}^{-1}(\boldsymbol{\theta}_{r})  =\frac{1}{\mathbf{g}^{\text{T}}_{r} \mathbf{A}\mathbf{g}_{r}} \text{adj}(\mathbf{J}(\boldsymbol{\theta}_{r}))
\end{equation}
where $\mathbf{A}=\frac{1}{2}\mathbf{f}\mathbf{t}^{\text{T}} +\frac{1}{2}\mathbf{t}\mathbf{f}^{\text{T}} -\mathbf{k}\mathbf{k}^{\text{T}}$.
As a result, we have the \ac{CRB} for the y-position coordinate $p_{\text{y}}$ as
\begin{equation}\label{eq:CRLBy}
\begin{split}
\left[\text{cov}\{\hat{\mathbf{p}}-\mathbf{p}\}\right]_{1,1} 
& = \sum_{r=1}^{\text{R}} \left(\frac{\sum_{r=1}^{\text{R}} k^{2}_{r}-k_{r}\sum_{r=1}^{\text{R}} k_{r} }{\text{R} \sum_{r=1}^{\text{R}} k^{2}_{r}-\left( \sum_{r=1}^{\text{R}} k_{r} \right)^{2}}\right)^{2} c_{r},
\end{split}
\end{equation}
and the \ac{CRB} for the x-position coordinate $p_{\text{x}}$ as
\begin{equation}\label{eq:CRLBx}
\begin{split}
\left[\text{cov}\{\hat{\mathbf{p}}-\mathbf{p}\}\right]_{2,2}
& = \sum_{r=1}^{\text{R}} \left(\frac{\sum_{r=1}^{\text{R}} k_{r}-\text{R}k_{r}}{\text{R} \sum_{r=1}^{\text{R}} k^{2}_{r}-\left( \sum_{r=1}^{\text{R}} k_{r} \right)^{2}} \right)^{2} c_{r}.
\end{split}
\end{equation}

\section{Numerical Experiment}
\begin{figure*}[!t]
\centering
\subfloat[Simulation scenario.]{\includegraphics[width=0.675\columnwidth]{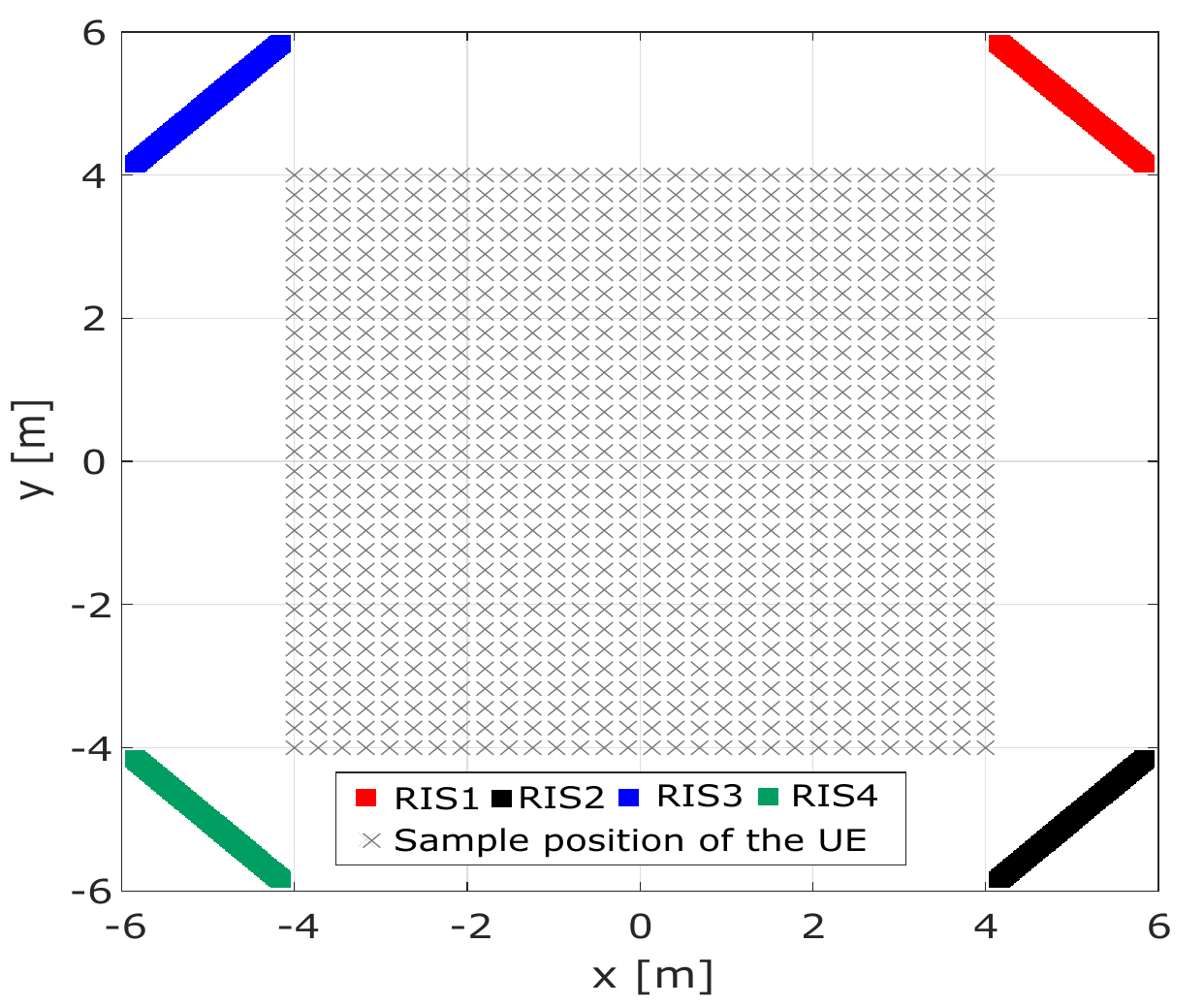}
\label{fig1}}
\hfil
\subfloat[Four RISs work randomly.]{\includegraphics[width=0.675\columnwidth]{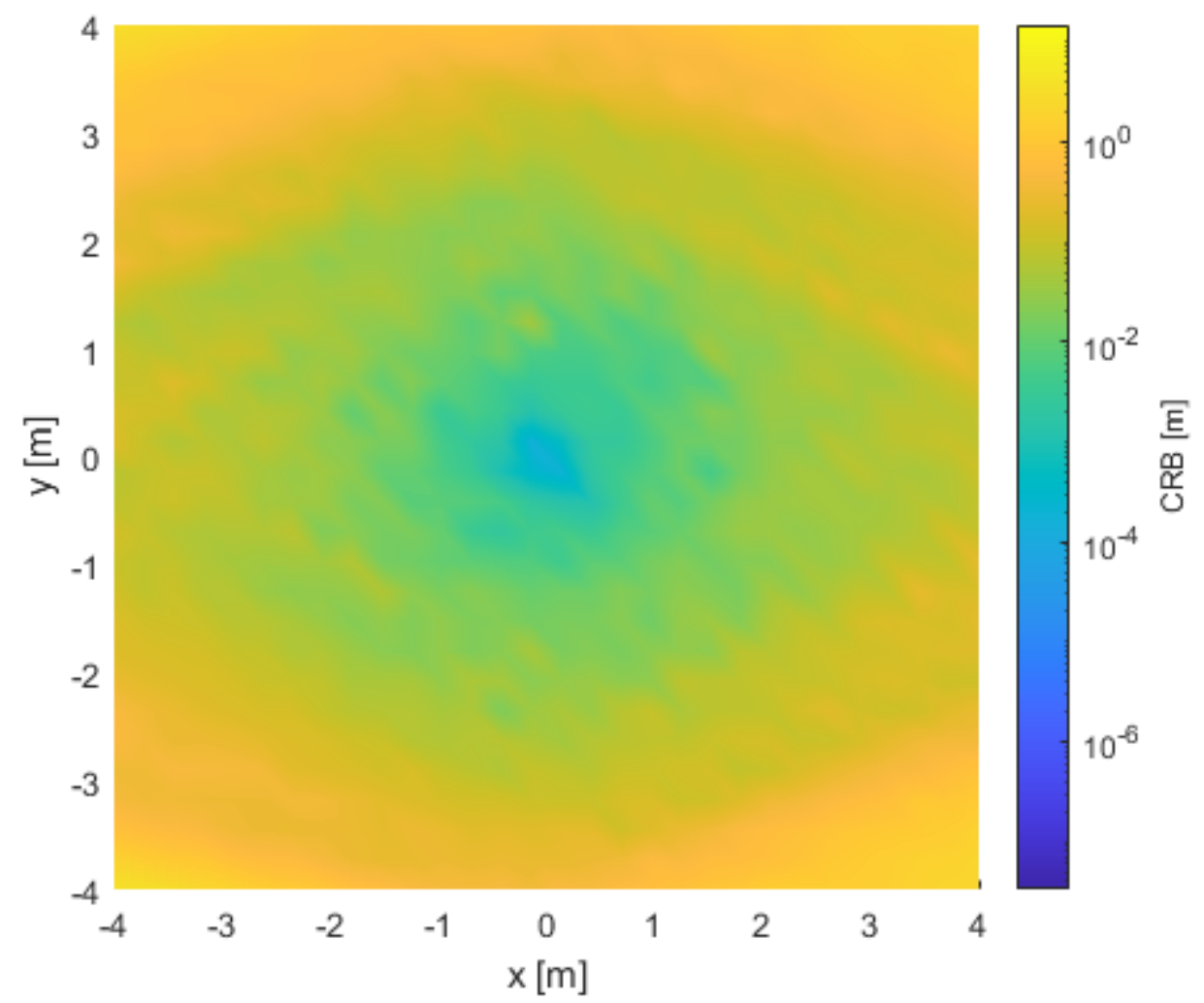}
\label{fig2}}
\hfil
\subfloat[Four RISs work like mirrors. ]{\includegraphics[width=0.675\columnwidth]{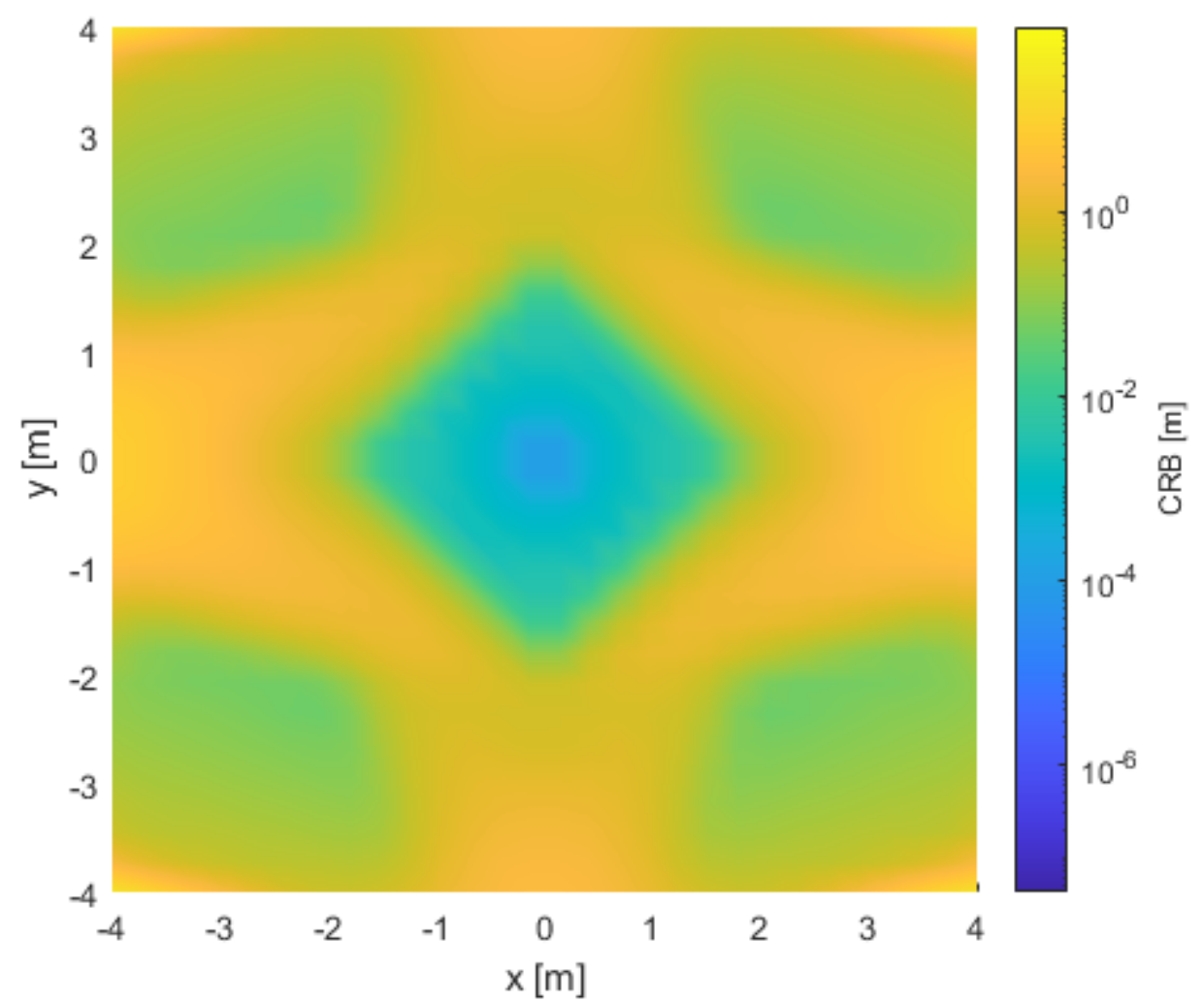}
\label{fig3}}
\vfil
\subfloat[Only RIS\,1,3 work randomly.]{\includegraphics[width=0.675\columnwidth]{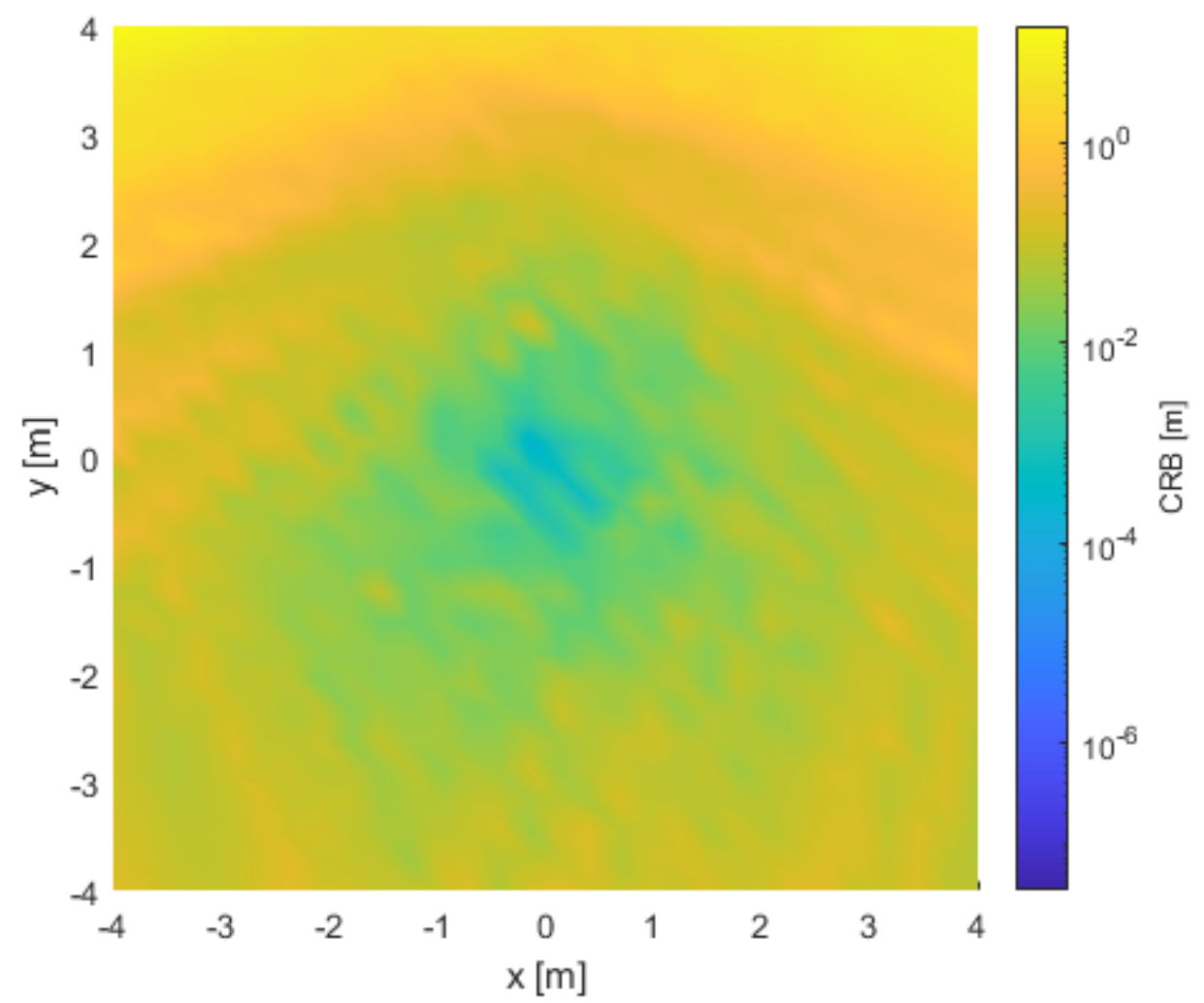}
\label{fig4}}
\hfil
\subfloat[Only RIS\,1,2 work randomly.]{\includegraphics[width=0.675\columnwidth]{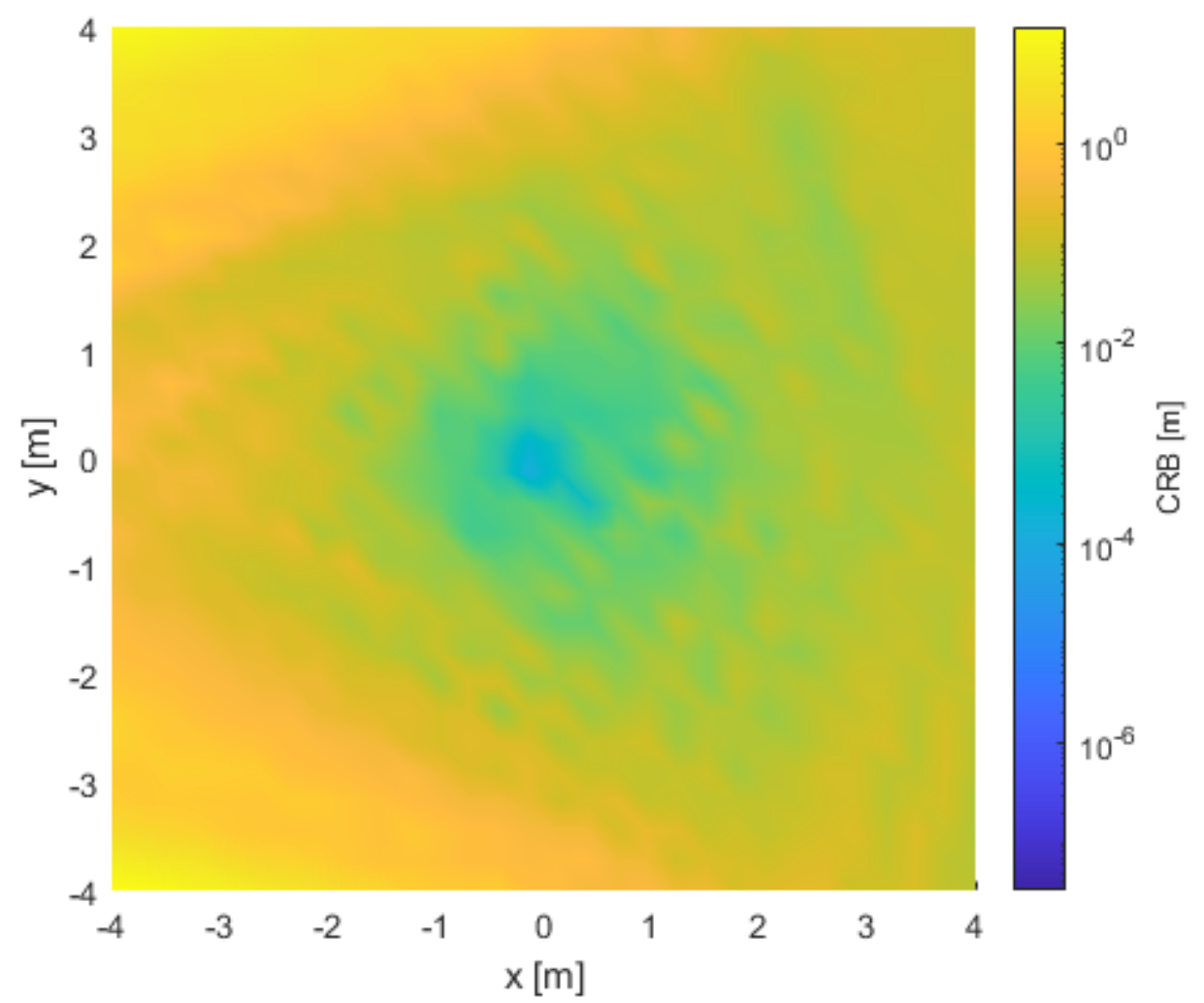}
\label{fig5}}
\hfil
\subfloat[Only RIS\,2,3 work randomly.]{\includegraphics[width=0.675\columnwidth]{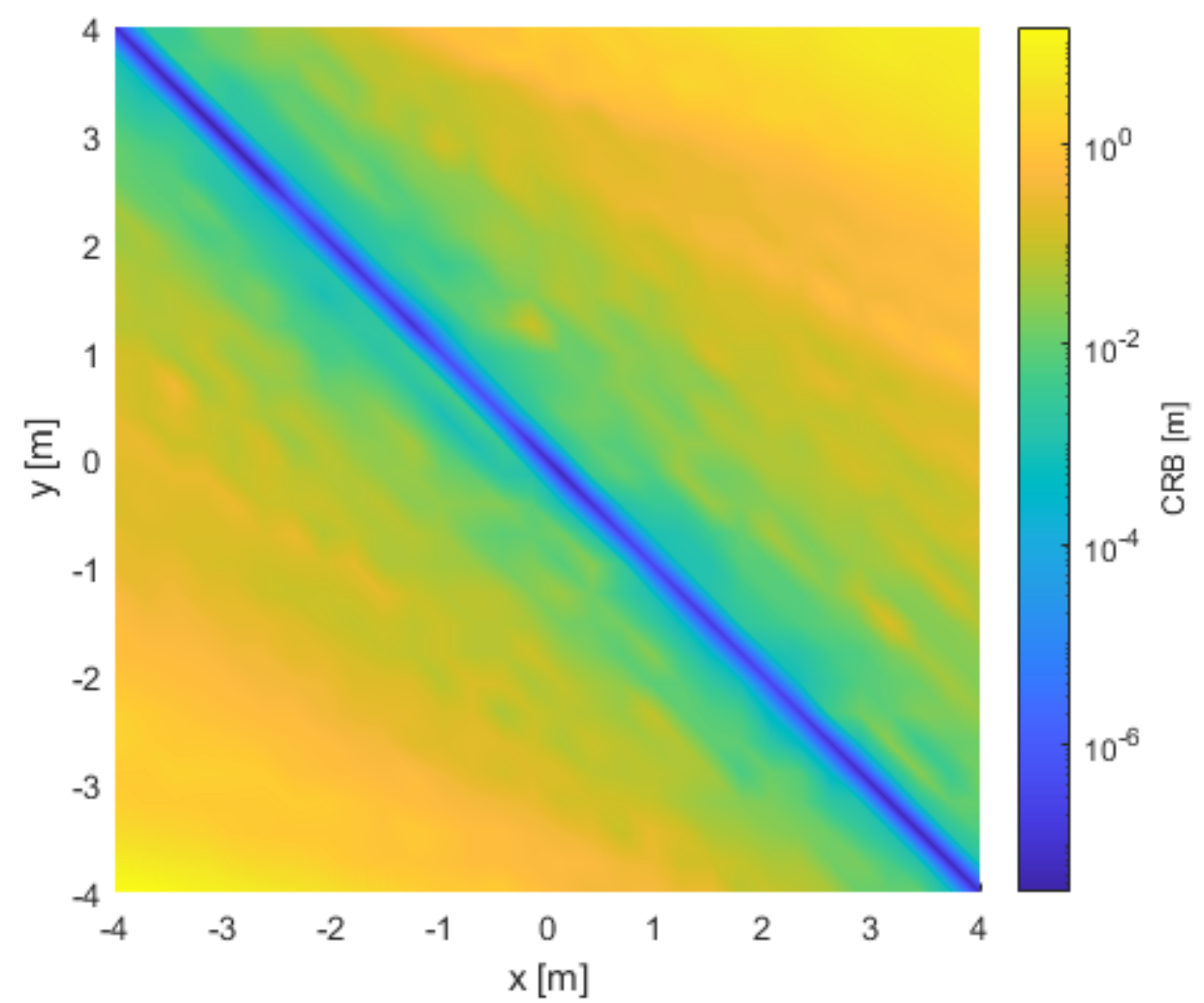}
\label{fig6}}
\vfil
\subfloat[Only RIS\,1,3 work like mirrors.]{\includegraphics[width=0.675\columnwidth]{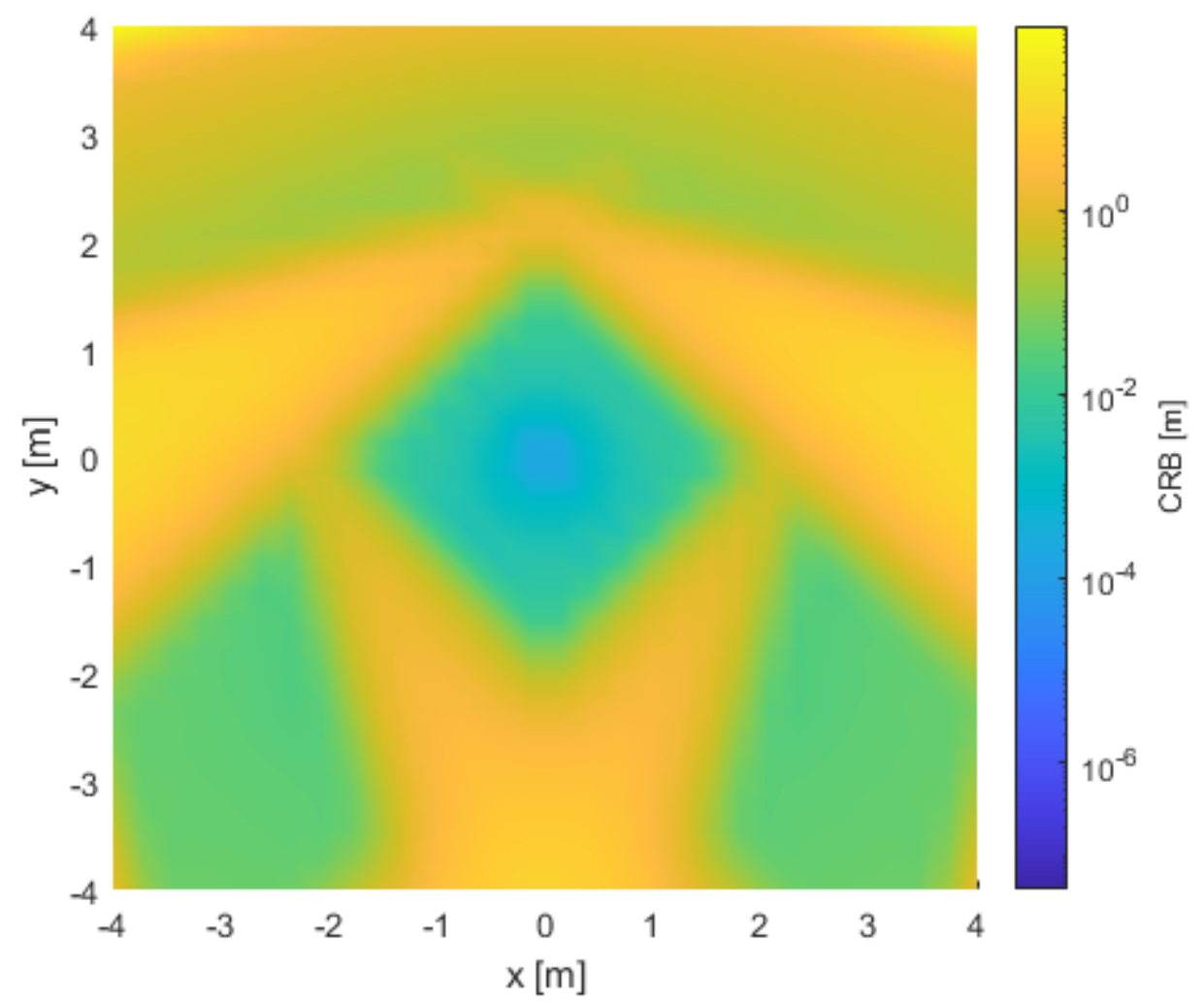}
\label{fig7}}
\hfil
\subfloat[Only RIS\,1,2 work like mirrors.]{\includegraphics[width=0.675\columnwidth]{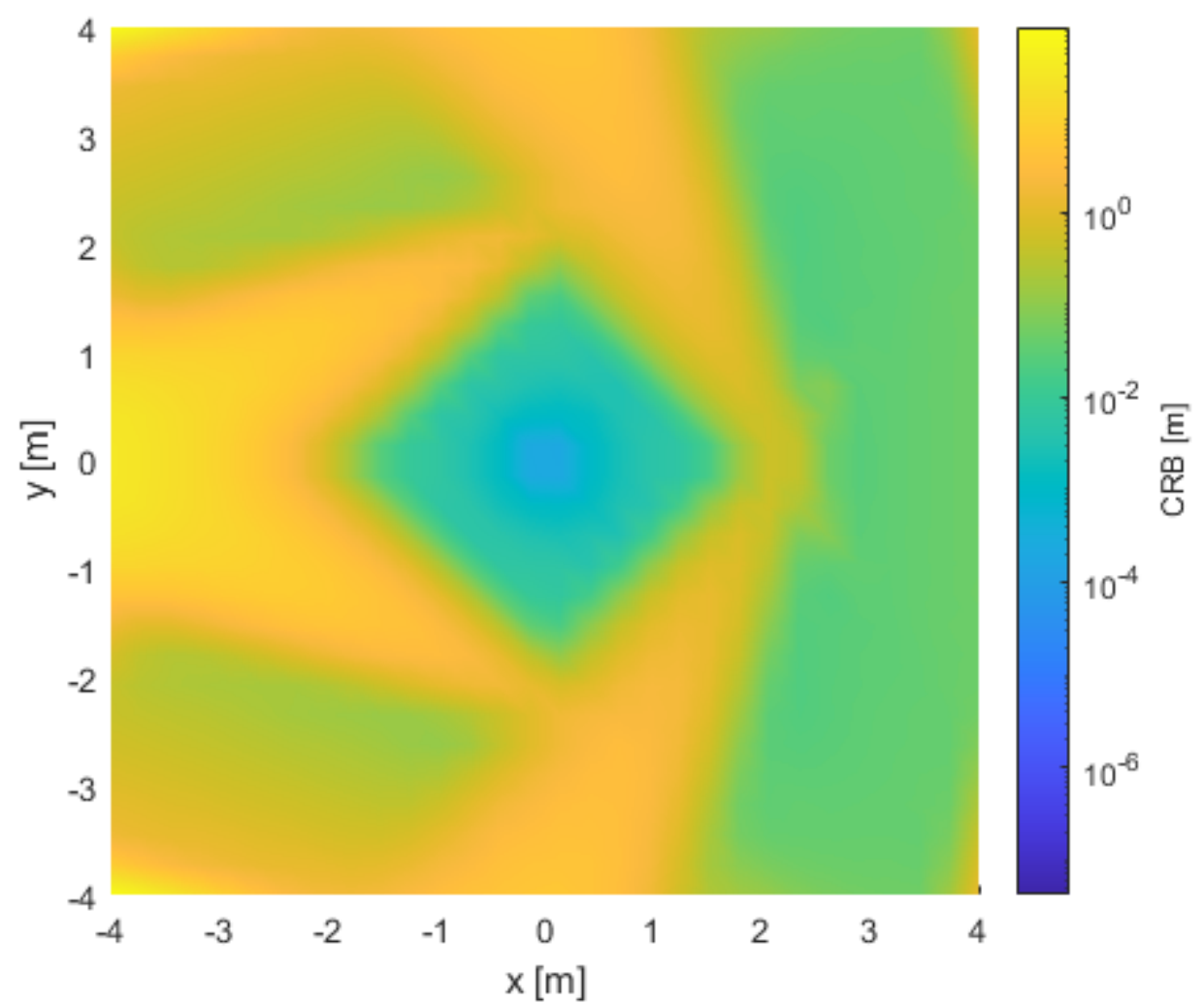}
\label{fig8}}
\hfil
\subfloat[Only RIS\,2,3 work like mirrors.]{\includegraphics[width=0.675\columnwidth]{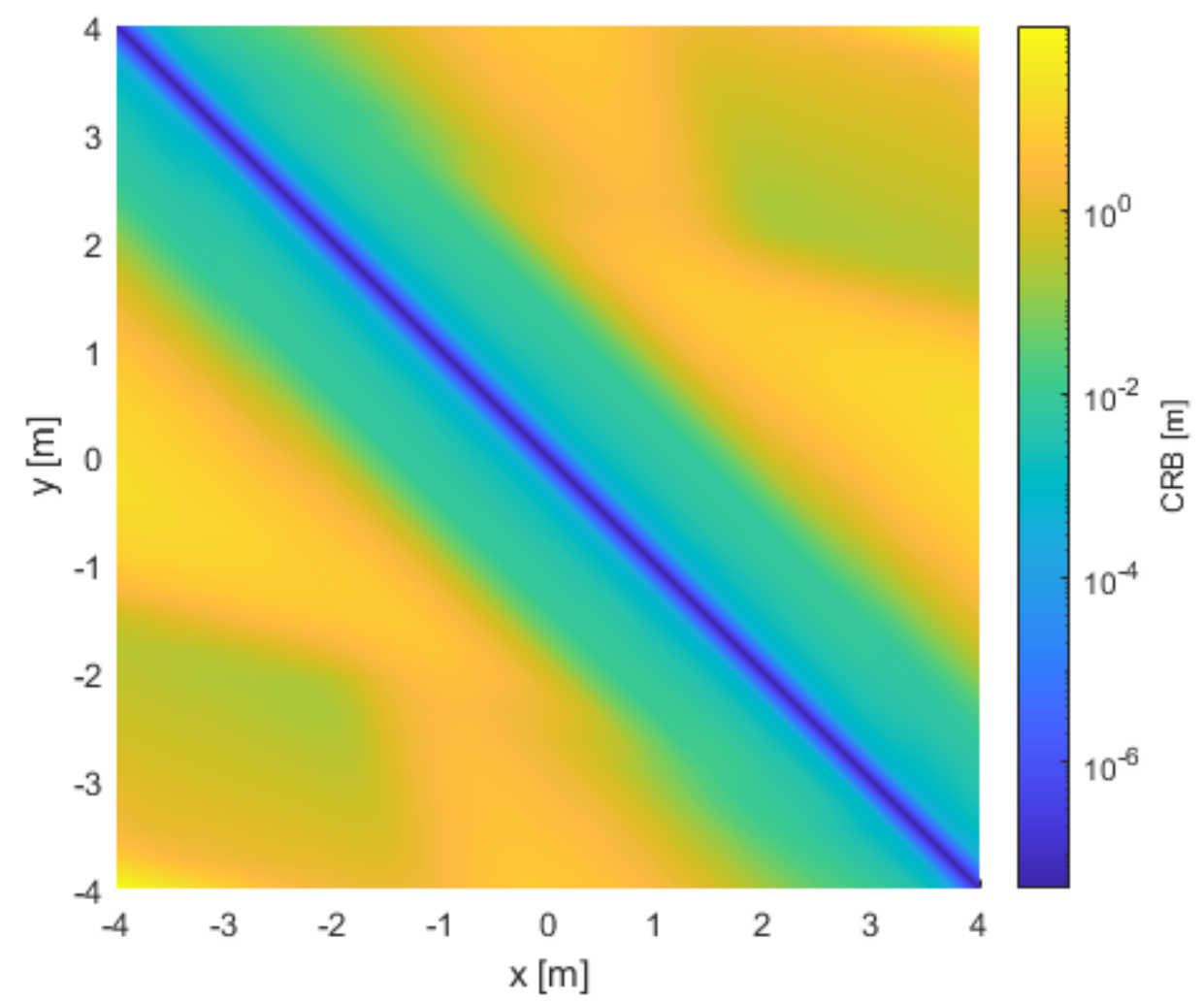}
\label{fig9}}
\caption{Simulated \ac{CRB} of $p_{\text{y}}$ under different conditions. In \cref{fig2,fig3} all the RISs are used for positioning. In \cref{fig4,fig5,fig6,fig7,fig8,fig9} only two RISs are used. Different system configurations can significantly affect the \ac{CRB} for various sample positions.}
\label{fig:sim}
\end{figure*}

In this section, we provide numerical experiment to illustrate the performance achievable by the considered positioning system. The scenario is configured as follows:
\begin{itemize}
\item Four \ac{ULA} \acp{RIS}: Each \ac{RIS} contains 100 pixels, arranged with half-wavelength spacing at a frequency of 25\,GHz. The location of the \acp{RIS} are as \cref{fig1} shows.
\item Four single-antenna \acp{BS}, each transmitting to a corresponding \ac{RIS}. Specifically, \ac{BS}\,1 (for \ac{RIS}\,1) is located at (251 m, 251 m), \ac{BS}\,2 (for \ac{RIS}\,2) at (251 m, -251 m), \ac{BS}\,3 (for \ac{RIS}\,3) at (-251 m, 251 m), and \ac{BS}\,4 (for \ac{RIS}\,4) at (-251 m, -251 m).
\item The transmitted power is uniformly distributed across both frequency and time. The \ac{SNR} at the BSs side is 47\,dB.
\item We consider a grid of $30\times30$ samples of the \ac{UE} position as \cref{fig1} shows. The \ac{UE} is moving at a speed of 10\,m/s
\item The transmission frequency band ranges from 24.5\,GHz to 25.5\,GHz, with the number of frequency samples $\mathrm{F}=201$. The transmission time span is 0.025 seconds, with a total of $\mathrm{T}=100$ time samples. 
\item The \acp{RIS} work either randomly (meaning that the scattering coefficients take values of either -1 or 1 at random) or like mirrors (all \ac{RIS} pixels have their scattering coefficients set to 1).
\end{itemize}
In \cref{fig:sim}, we present the \ac{CRB} of  $p_{\text{y}}$ under various conditions\footnote{In our simulation setup, the symmetrical nature of the scenario causes $p_{\text{y}}$ and $p_{\text{x}}$ to behave equivalently.}. 

\subsection{The impact of configuration of RISs}
A comparison of \cref{fig2} and \cref{fig3}, \cref{fig4} and \cref{fig7}, \cref{fig5} and \cref{fig8}, and \cref{fig6} and \cref{fig9} reveals that RIS configuration significantly affects the CRB heatmap, even when the number and placement of RIS elements are unchanged. For instance, a mirror-like RIS yields a lower CRB at the center compared to a random configuration, but increases CRB in surrounding regions. This highlights how RIS configuration critically shapes the spatial distribution of CRB, reflecting a trade-off between central and peripheral performance. In our future work, we plan to utilize this insight to optimize the \ac{RIS} configurations.

\subsection{The impact of number of RISs}
Analysis of \cref{fig7} and \cref{fig8} reveals that with only two RISs, some regions in the heatmap show notably higher CRB values. Introducing another two RISs (\cref{fig3}) lowers the maximum CRB yet raises it in areas previously exhibiting low values, particularly near the RISs. This highlights the need for further refinement of the position estimation method, in particular, replacing the \ac{LS} estimate with a more elaborate estimate such as \ac{ML} estimation or Bayesian estimation.

\subsection{The impact of locations of RISs}
As shown in Figures  \cref{fig7} and \cref{fig8}, even with only two RISs, their spatial configuration significantly influences the CRB distribution. Notably, \cref{fig8} features a region with markedly lower CRB values than \cref{fig7}, underlining the importance of optimal RIS placement. However, this finding also could suggest that the theoretical CRB may not fully capture real-world estimation errors, indicating a possible gap between predicted and actual performance. Consequently, further investigation is needed to clarify how RIS positioning, CRB, and estimation accuracy interact.

\section{Conclusion and Outlook}
The results in this work underscore the potential of the proposed scheme. In future work, we plan to explore more realistic scenarios (such as multiple antenna \ac{BS} and the impact of wideband transmission\cite{huang2024single}), refine the methodology, and establish a comprehensive solution for practical applications.

\section*{Acknowledgment}
This research was funded in whole or in part by the Austrian Science Fund (FWF) 10.55776/PAT4490824.

\bibliographystyle{IEEEtran}
\bibliography{references}

\end{document}